\documentclass{nature_modified_OP}

\usepackage{graphicx} 
\usepackage{caption}
\renewcommand{\figurename}{Fig.}

\usepackage{amsfonts}
\usepackage{mathtools}
\usepackage{amsmath}
\usepackage{array}

\usepackage{multirow}
\usepackage{amssymb}
\usepackage{lineno}

\usepackage{xcolor}
\usepackage{soul}
 
\usepackage{graphicx}  
\usepackage{xcolor}
\usepackage{newfloat} 
\DeclareFloatingEnvironment[name={Extended Data Figure}]{suppfigure}

\title{Decoding the human brain tissue response to radiofrequency excitation using a biophysical-model-free deep MRI on a chip framework}

\author{Dinor Nagar$^1$, Moritz Zaiss$^{2,3}$, Or Perlman$^{4,5*}$}

\begin{document}

\maketitle

\begin{affiliations}
 \item School of Electrical Engineering, Tel Aviv University, Tel Aviv, Israel
 \item Institute of Neuroradiology, Friedrich-Alexander Universität Erlangen-Nürnberg (FAU), University Hospital Erlangen, Erlangen, Germany
 \item Department of Artificial Intelligence in Biomedical Engineering, Friedrich-Alexander Universität Erlangen-Nürnberg, Erlangen, Germany
 \item Department of Biomedical Engineering, Tel Aviv University, Tel Aviv, Israel 
 \item Sagol School of Neuroscience, Tel Aviv University, Tel Aviv, Israel

\end{affiliations}

\noindent{$^*$Correspondence to: Or Perlman, Department of Biomedical Engineering and Sagol School of Neuroscience, Tel Aviv University, Multidisciplinary Research Building, Room 410, Tel Aviv, 6997801, Israel. Email: orperlman@tauex.tau.ac.il}\\

\newpage

\begin{abstract}
\textbf{Abstract}\\
Magnetic resonance imaging (MRI) relies on radiofrequency (RF) excitation of proton spin. Clinical diagnosis requires a comprehensive collation of biophysical data via multiple MRI contrasts, 
 acquired using a series of RF sequences that lead to lengthy examinations. Here, we developed a vision transformer-based framework that captures the spatiotemporal magnetic signal evolution and decodes the brain tissue response to RF excitation, constituting an MRI on a chip. Following a per-subject rapid calibration scan (28.2 s), a wide variety of image contrasts including fully quantitative molecular, water relaxation, and magnetic field  maps can be generated automatically. The method was validated across healthy subjects and a cancer patient in two different imaging sites, and proved to be 94\% faster than alternative protocols. The deep MRI on a chip (DeepMonC) framework may reveal the molecular composition of the human brain tissue in a wide range of pathologies, while offering clinically attractive scan times. 
\end{abstract} 

\noindent Magnetic resonance imaging (MRI) is among the most powerful diagnostic tools in present-day clinical healthcare\cite{van2019value}. One compelling advantage is its wide versatility, which allows a single imaging modality to acquire a wide variety of biophysical information\cite{yousaf2018advances}. This is a consequence of the ability to program MRI scans for emphasizing a particular tissue property of interest. Specifically, a series of radiofrequency (RF) pulses are designed and applied to initiate a cascade of interactions with the tissue proton spins. The particular waveform, duration, power, and frequency of each RF pulse, as well as the characteristics of the entire RF train ensemble affect the resulting contrast,  which can be customized to detect microstructure, water content, cellularity, blood flow, molecular composition, and even functional characteristics\cite{bernstein2004handbook}. 

As a single MRI pulse sequence is generally not enough to determine the tissue state and make a diagnosis with sufficient certainty, standard clinical MRI exams involve the serial acquisition of multiple pulse sequences\cite{ellingson2015consensus}. For example, brain cancer MRI protocols typically comprise T$_1$-weighted, T$_2$-weighted, fluid-attenuated inversion recovery (FLAIR) and diffusion, and may also require perfusion, and MR-spectroscopy imaging\cite{kaufmann2020consensus}. While multi-sequence acquisition provides rich information, it also requires excessively long examination times (e.g., 20-60 min) that increase patient discomfort, healthcare costs, and waiting lines\cite{edelstein2010mri}. Moreover, the image contrast  depends not only on the acquisition parameters, but also on the particular tissue characteristics, which may be highly variable across subjects. For example, while a given protocol may be able to differentiate multiple tumor components in one patient, it may be insufficient for another individual, and may require fine-tuning of the RF pulses (e.g., choosing a different flip angle, saturation pulse power, etc.). As radiological image analysis is commonly performed "offline", namely, after the acquisition is completed and the subject has left, rescheduling a subject scan for re-imaging with a modified acquisition protocol is either impractical or results in increased costs and prolonged waiting lines\cite{sreekumari2019deep}.

In recent years, the need to enhance the biochemical information portfolio provided by MRI has prompted the search for new contrast options. Saturation transfer (ST) MRI represents one such promising option\cite{van2018magnetization}, due to the RF-tunable sensitivity for various molecular properties, such as
mobile protein and peptide volume fractions, intracellular pH, and glutamate concentration\cite{liu2013nuts}. ST-MRI has shown promise for a variety of clinical applications\cite{jones2018clinical}, such as tumor detection and grading\cite{zhou2011differentiation, zhou2019apt}, early stroke characterization\cite{zhou2003using, sun2007detection}, neurodegenerative disorder imaging\cite{cai2012magnetic, cember2023glutamate}, and kidney disease monitoring\cite{stabinska2023cest, longo2017noninvasive}. However, the integration of ST-MRI into clinical practice has been slow and limited, because of the relatively long scan times required. Moreover, as each ST target compound is characterized by a distinct proton exchange rate, a separate pulse sequence must be acquired for each application of interest, thereby rendering multi-contrast ST imaging even less practical. 

Quantitative imaging of biophysical tissue properties offers improved  reproducibility, sensitivity, and consistency across sites and scanners, compared to contrast-weighted imaging\cite{seiberlich2020quantitative}. In this context, imaging techniques that combine biophysical (differential-equation-based) models with artificial intelligence (AI) have recently been suggested to be useful for accelerating water-pool relaxometry\cite{ma2013magnetic, fujita2023mr, gaur2023magnetic, cohen2018mr, fyrdahl2022magnetic} or quantitative ST-MRI acquisition and reconstruction\cite{perlman2023mr, perlman2022quantitative, perlman2022end, cohen2023cest, nagar2023dynamic, kang2022learning, vladimirov2024quantitative}. Unfortunately, the complexity of the multi-proton-pool in-vivo environment and the challenges in accurately modeling the large number of free tissue parameters, limit the efficacy of this approach for molecular MRI. This leads to substantial variability between the biophysical values reported by various groups (each incorporating different model assumptions)\cite{weigand2023accelerated, heo2019quantifying, carradus2023measuring} or to increased acquisition times, as water pool and magnetic field parameters may need to be separately estimated via additional pulse sequences, to reduce the model complexity\cite{perlman2022quantitative, kim2020deep}.

Here, we describe the development of a biophysical-model-free deep learning framework (\textbf{Fig. 1}), which can provide rich biological information in-vivo, while circumventing the need for lengthy multi-pulse-sequence MRI acquisition (requires only a 28.2 second-long calibration scan). The framework is able to capture the spatiotemporal magnetic signal evolution dynamics in living humans and decode the brain tissue response to RF excitation, constituting a deep MRI on a chip (DeepMonC). When employed on unseen subjects, pathology, and scanner models at a different imaging site from where the training set was obtained, DeepMonC was able to emulate the spin evolution dynamics accurately, and generate a variety of new image contrasts as well as fully quantitative molecular, water relaxation, and magnetic field maps.   

\section*{Results}
\section*{DeepMonC Framework}
The DeepMonC core module (\textbf{Fig. 1a}) was designed to capture the spatiotemporal dynamics of MRI signal propagation as a response to RF excitation, and enable the generation of on-demand image contrast. The system includes a vision transformer\cite{hatamizadeh2022unetr, dosovitskiy2020image} with a dual-domain input, comprised of RF excitation information and real-world tissue response image counterparts. An extension module was also designed, which quantifies six biophysical tissue parameters across the entire 3D brain, without the need for any additional input. 

The \textit{core module} inputs are a sequence of \textit{m}=6 non-steady-state MRI calibration images and an RF excitation parameter tensor (\textbf{Fig. 1a}). The tensor includes two concatenated parts: the acquisition parameters used for obtaining the calibration images and the desired on-demand parameters for the subsequent image output. Separate embeddings for the real-image-data and the physical RF properties are then learned using a vision transformer and a fully connected layer, respectively. The \textit{quantification module}, involves a transfer learning strategy where the core module weights are plugged-in, the last layer is removed, and there is augmentation of two new convolutional layers. Ground truth reference data are then used to instigate quantification-oriented learning (\textbf{Fig. 1b}). 

The DeepMonc framework was trained using 3,118,692 image and acquisition parameter pairs from 9 healthy human volunteers, scanned at a single imaging site (Tel Aviv University) on a 3T MRI (Prisma, Siemens Healthineers) equipped with a 64-channel coil. The framework was then tested using 30,324  image and acquisition parameter pairs obtained from 4 other subjects representing three challenging datasets: (i) Two healthy subjects not used for training (scanned at the same site). (ii) A brain cancer patient scanned at a different imaging site (Erlangen University Hospital). (iii) A healthy volunteer scanned using different hardware and MRI model at a different imaging site (Erlangen University Hospital, Trio MRI with a 32 channel coil).

\section*{Biophysical-model-free prediction of the tissue response to RF excitation}
The core module was validated for generating on-demand molecular (semisolid MT and amide proton CEST-weighted) images. The full reference imaging protocol consisted of 30 pseudo-random RF excitations (\textbf{Supporting Information Fig. 1})\cite{perlman2022quantitative}. The first six images were used for per-subject calibration, followed by DeepMonC predictions of the multi-contrast images associated with the next six response images (\textbf{Fig. 1a}). 
A representative example of the DeepMonC output compared to the ground truth for each of the validation datasets is shown in \textbf{Fig. 2} and whole-brain 3D reconstruction output is provided as \textbf{Supporting Information Movies M1} (semisolid MT) and \textbf{M2} (amide). An excellent visual, perceptive, and pixelwise similarity was obtained between DeepMonC output and  ground truth. This is reflected by a  structural similarity index measure (SSIM) $>$ 0.96, peak signal-to-noise ratio (PSNR) $>$ 36, and normalized mean-square error (NRMSE) $<$ 3\% (\textbf{Table 1}).

To evaluate the ability to generate an up to 4-times longer output compared to the input, the process was continued recursively, until the entire 30-long sequence was predicted based on the first six calibration images (\textbf{Supporting Information Movies M3} (semisolid MT) and \textbf{M4} (amide)). Although there were some errors in the last six images, the overall performance remained high, with a structural similarity index measure (SSIM) $>$ 0.94, peak signal-to-noise ratio (PSNR) $>$ 32, and normalized mean-square error (NRMSE) $<$ 3.7\% (\textbf{Table 1}). The inference times for reconstructing whole brain 6 or 24 unseen image contrasts were 7.674 s and 10.896 s, respectively, when using an Nvidia RTX 3060 GPU, and 9.495 s and 19.55 s, respectively, when using a desktop CPU (Intel I9-12900F).

\section*{Rapid quantification of biophysical tissue parameters}
The quantification module was trained to receive the exact same input as the core module, and then produce six parameter maps: the semisolid MT proton volume fraction (f$_{ss}$) and exchange rate (k$_{ssw}$), water pool longitudinal (T$_1$) and transverse (T$_2$) relaxation times, and the static (B$_0$) and transmit (B$_1$) magnetic fields. The DeepMonC reconstructed paramater maps were visually, perceptually, and quantitatively similar to the ground truth reference (\textbf{Fig. 3-5 panels a,b} and \textbf{Supporting Information Figure S2}). The reconstruction performance was highest for the test subject scanned by the same scanner used for training (SSIM = 0.919$\pm$0.024;  PSNR = 30.197$\pm$1.808; NRMSE = 0.049$\pm$0.008), followed by the cancer patient (unseen pathology at an unseen imaging site: SSIM = 0.884$\pm0.024$; PSNR = 26.349$\pm$1.246; NRMSE = 0.059$\pm$0.007), and the unseen subject scanned using unseen hardware at an unseen imaging site (SSIM = 0.811$\pm$0.044;  PSNR = 24.186$\pm$1.523; NRMSE = 0.076$\pm$0.011).  

The magnetic field maps reconstructed by DeepMonc exhibited improved homogeneity compared to their  ground-truth counterparts (\textbf{Fig. 3,4,5 panels a and b}). This enabled successful artifact removal from the semisolid MT proton volume fraction and exchange rate maps, which are known to be sensitive to B$_0$ and B$_1$ inhomogeneity\cite{windschuh2015correction, samson2006simple, schuenke2017simultaneous} (white arrows in \textbf{Fig. 3}, and \textbf{Fig. 5}).

To analyze the contribution of the decoded tissue response information, captured by DeepMonc core module, to the quantification task performance, a comparison with standard supervised learning was performed. The same quantification architecture (\textbf{Fig. 1b}) was trained to receive the exact same inputs, and then output the same six quantitative biophysical parameter maps, but without employing the pre-trained DeepMonC weights (learnt by the core module, \textbf{Fig. 1a}). This standard supervised learning routine yielded parameter maps with a markedly lower resemblance to the ground truth (\textbf{Fig. 3,4,5 panel c}). The deterioration in output was accompanied by a statistically significant lower SSIM (0.805$\pm$0.057, 0.778$\pm$0.062, 0.725$\pm$0.066, for the unseen subject, pathology, and hardware datasets, respectively, p$<$0.0001, n=68 image pairs) and PSNR (25.733$\pm$1.473, 23.546$\pm$1.428, 22.614$<$1.342, for the three datasets, respectively,  p$<$0.0001, n=68 image pairs), and a higher NRMSE (0.0842$\pm$0.0125, 0.0843$\pm$0.0128, 0.092$\pm$0.012 for the three datasets, respectively, p$<$0.0001, n=68 image pairs, \textbf{Fig. 3,4,5 panel d}). The inference time required for reconstructing whole brain quantitative images was 6.751 s or 9.822 s  when using an Nvidia RTX 3060 GPU or a desktop CPU (Intel I9-12900F), respectively.

\section*{Discussion}
The past few decades have seen increased reliance on MRI for clinical diagnosis\cite{smith2019trends}. In parallel, this has required the introduction of new contrast mechanisms and dedicated pulse sequences\cite{ward2000new, wang2017clinical, wu2016overview, jones2018clinical, van2011chemical, marrale2016physics, duhamel2019validating}. While offering biological insights and improved diagnosis certainty, the integration of these sequences into routine MRI examinations exacerbates the already lengthy overall scan times. Here, we describe the development of a deep-learning-based framework that can rapidly decode the human brain tissue response to RF excitation. The system generates a variety of on-demand image contrasts in silico that faithfully recapitulate their physical in-vivo counterparts (hence, termed a deep MRI on a chip). 

The target contrasts requested from DeepMonC were associated with RF parameters extrapolated beyond the range of the training parameters, thereby representing a highly challenging task (\textbf{Supporting information Fig. 1}). Nevertheless, an excellent agreement between the generated and ground-truth image-sets was obtained (\textbf{Fig. 2} and \textbf{Table 1}). The dependence of DeepMonC on the particular set of calibration images used and the desired output contrast was assessed on 18 different input-output pairs (\textbf{Supporting Information Figure S3}). Despite some variability, a satisfactory reconstruction was obtained in all cases (SSIM $>$ 0.96, PSNR $>$ 36, NRMSE $>$ 2\%). Importantly, DeepMonC was able to overcome  unknown initial conditions, as all calibration image-set combinations but one (image indices 1-6, \textbf{Supporting Information Figure S3}) were acquired following an incomplete magnetization recovery.

The core module architecture was designed for image translation of \textit{m}-to-\textit{m} size \textbf{(Fig. 1a, illustrated for m=6)}. Nevertheless, it can be recursively applied (by using the model's output as the next input for generating another set of \textit{m} images), and maintains an attractive performance, for up to \textit{m}-to-3\textit{m} translations (\textbf{Supporting Information Movies M3} and \textbf{M4}). Although some errors were visually observed when attempting \textit{m}-to-4\textit{m} translation (in the last \textit{m}=6 images), additional training with longer acquisition protocols could further improve this performance. 

The excellent on-demand contrast generation performance exhibited by DeepMonC (\textbf{Table 1}) can be attributed to two key factors: (1) The introduction of explicit (and varied) acquisition parameter descriptors into the training procedure; this information is traditionally overlooked and hidden from MR-related neural networks\cite{heckel2024deep, chen2022ai}. (2) The incorporation of visual transformers as the learning strategy. These enable the system to address the double sequential nature of the image data obtained from both the 3D spatial domain and the temporal (spin-history) domain. Visual transformers, with their effective attention mechanism, are not only capable of capturing long-range data dependencies but can also understand global image context, alleviate noise, and adapt to various translational tasks\cite{dosovitskiy2020image, khan2022transformers}.

Contrast-weighted imaging is the prevalent acquisition mode in clinical MRI. However, it has become increasingly clear that \textit{quantitative} extraction of biophysical tissue parameters may offer improved sensitivity, specificity, and reproducibility\cite{buonincontri2021three, panda2019repeatability, yankeelov2011quantitative}. By harnessing the decoded brain tissue response to RF excitation, the DeepMonC framework was further leveraged to simultaneously map six quantitative parameters \textbf{(Fig. 3-5}), spanning three different biophysical realms, namely water relaxation, semisolid macromolecule proton exchange, and magnetic field homogeneity. The results provide an excellent agreement with the ground truth (\textbf{Fig. 3-5d}, \textbf{Supporting Information Fig. S2}), as well as an inherent ability to mitigate artifacts (white arrows in \textbf{Fig. 3} and \textbf{Fig. 5}). Specifically, the B$_0$ and B$_1$ maps generated by DeepMonC exhibit better homogeneity than the reference ground truth. This thereby represents a practical explanation for the successful reduction of hardware/in-homogeneity related noises around the sinuses/eyes and at the air-tissue interfaces.

Importantly, the rich whole-brain information provided by DeepMonc was reconstructed in only 6.8 seconds, following a non-steady state rapid acquisition using a \textit{single} pulse sequence of 28.2 s. This represents a 94\% acceleration compared to the state of the art ground-truth reference (acquired in 8.5 min, \textbf{Fig. 1b}). Interestingly, the quantification task results were even less sensitive to the particular pulse sequence  used for acquiring the calibration images (\textbf{Supporting Information Figure S4}) than the on-demand contrast generation task (\textbf{Supporting Information Figure S3}). 

The success of the quantification module is directly associated with the reliance on DeepMonC's core pre-training, which generates a comprehensive understanding of the RF-to-tissue relations. This is supported by the statistically significant higher performance obtained by the quantification module compared to the vanilla use of DeepMonC (untrained) architecture (\textbf{Fig. 3-5 panels c,d}, n=68 image slices, p$<$0.0001).

The generalization of DeepMonC predictions was assessed on three datasets, each representing a different challenge.  Overall, there proved to be compelling evidence for generalization, with a faithful representation of the the RF-to-tissue interface, with a satisfactory image reconstruction obtained in all cases. It should however be noted that, as expected, the parameter quantification of the unseen subject scanned at the same site and scanner used for training, yielded the best results. The cancer patient scanned at a different image site yielded the next best performance (only healthy volunteers were used for training), followed by the healthy subject scanned using a different scanner model and hardware at a different imaging site (\textbf{Fig. 3-5d}, \textbf{Supporting Information Fig. S4}). When assessing the on-demand contrast generation task performance, the differences between the various test-sets were much less discernible, with mostly subtle variations in the reconstruction metrics (\textbf{Table 1}). In the future, additional training using subjects scanned on other scanner models and across various pathologies could further boost the framework performance.

Saturation transfer (encompassing both CEST and semisolid MT) is the dominant biophysical mechanism involved in the on-demand contrast generation task. This was chosen as a representative emerging imaging approach that is the focus of much interest from across the medical community\cite{jones2018clinical, kim2016emerging, zhou2022review, jabehdar2023chemical, van2021apt, rivlin2023metabolic, armbruster2024personalized}.  Nevertheless, the same conceptual framework could potentially be applied for generating on-demand diffusion, perfusion, relaxation, susceptibility, and other contrast-weighted images, given that a per-subject rapidly acquired data from the same general mechanism of interest is provided, alongside the matching acquisition parameters. Notably, a single pulse sequence may represent several biophysical properties, similarly to the way that ST-contrast weighted images are affected by the T$_1$, T$_2$, B$_0$, and B$_1$. Furthermore, while this work was focused on brain imaging, we expect that the same framework could be similarly utilized in other organ/tissues (after proper training). Finally, the ground-truth reference used for the quantification task was obtained via standard water proton relaxometry, magnetic field-mapping, and semisolid MT MRF. However, the same quantification module could seamlessly be trained using alternative reference modalities, such as 31P-imaging (for reconstructing intracellular pH maps)\cite{paech2023whole}, or even non-MR images (such as Ki-67 proliferation index histological images), thereby creating new cross-modality insights and opportunities.

In summary, we have developed and validated a computational framework that can learn the intricate mapping between the magnetic resonance RF irradiation domain and the subject-specific image domain. The method is biophysical-model-free and thus, unbiased by pre-existing parameter restrictions or assumptions. Given its ultra-fast on-demand contrast generation ability, we expect this approach to play an important role in the efforts to accelerate clinical MRI. 

\section*{Acknowledgments}
The authors thank Tony Stöcker and Rüdiger Stirnberg for their help with the 3D EPI readout. This project was funded by the European Union (ERC, BabyMagnet, project no. 101115639). Views and opinions expressed are however those of the authors only and do not necessarily reflect those of the European Union or the European Research Council. Neither the European Union nor the granting authority can be held responsible for them.

\section*{Author contributions}
Conceptualization: D.N., O.P., Deep learning methodology: D.N., O.P., MRI acquisition and reconstruction: M.Z., O.P, Writing, reviewing, and editing: D.N., M.Z., O.P., Supervision: O.P.

\section*{Competing interests}
D.N. and O.P applied for a patent related to the proposed framework.

\section*{Methods}

\subsection{Human subjects.}
Eleven healthy volunteers (five females/six males, with average age 25.5$\pm4.7$) were scanned at Tel Aviv University (TAU), using a 64-channel 3T MRI (Prisma, Siemens Healthineers). The research protocol was approved by the TAU Institutional Ethics Board (study no. 0007572-2) and the Chaim Sheba Medical Center Ethics Committee (0621-23-SMC). Two additional subjects were scanned at University Hospital Erlangen (FAU): a glioblastoma patient (World Health Organization grade IV, IDH mutation, and methylation of MGMT (O(6)-methylguanine-DNA methyltransferase) promoter), scanned using the same scanner model described above, and a healthy volunteer, imaged using a different 3T MRI model and coil system (Trio, Siemens Healthineers with a 32 channel coil). The research protocol was approved by the University Hospital Erlangen Institutional Review Board and Ethics Committee. All subjects gave written, informed consent before the study.

\subsection{MRI acquisition.} 
 Following scout image positioning and shimming, each subject was scanned using five different pulse sequences, all implemented using the Pulseq prototyping framework\cite{ravi2019pypulseq, layton2017pulseq} and the open-source Pulseq-CEST sequence standard\cite{herz2021pulseq}. Non-steady-state ST images of the amide and semisolid MT proton pools were acquired using two dedicated pulse sequences, as described previously\cite{perlman2022quantitative, weigand2023accelerated}. Each protocol employed a spin lock saturation train (13$\times$100 ms, 50\% duty-cycle), which varies the saturation pulse power between 0 and 4$\mu$T (detailed pattern available in \textbf{Supporting Information Figure S1}) to generate 30 contrast-weighted images. The saturation pulse frequency offset was fixed at 3.5 ppm for amide imaging\cite{perlman2020cest, cohen2018rapid
} or varied between 6 and 14 ppm for semisolid MT imaging\cite{perlman2022quantitative}. The saturation block was fused with a 3D centric reordered EPI readout module\cite{akbey2019whole, mueller2020whole}, which provided a 1.8 mm isotropic resolution across a whole-brain field of view. The echo time was 11 ms and the flip angle was set to 15$^{\circ}$. The same rapid readout module and hybrid pulseq-CEST framework were used to acquire additional B$_0$ and B$_1$ maps by using the WASABI method\cite{schuenke2017simultaneous}, and water T$_1$ and T$_2$ maps by using saturation recovery and multi-echo sequences, respectively. The total scan time per subject for all five protocols was 10.85 min (8.5 min for the quantitative reference set described in \textbf{Fig. 1b}).

\subsection{Data organization.}
The data from nine healthy subjects scanned at TAU was used for training and validation (hyper-parameter tuning) with a $\sim$80\%-20\% split. Each training sample was composed of an \textit{m}=6 image series, and an acquisition parameter tensor, which included the corresponding six saturation pulse power and frequency offset values utilized, as well as the parameters associated with the  subsequent \textit{m}-long image-series output (\textbf{Fig. 1a} and \textbf{Supporting Information Figure S1}). A 3D volume of maximum 144x144x144 voxel size was acquired for each subject. After all non-brain-containing slices were removed, 8-fold rotation-based data augmentation was performed. The core module was trained using 18 various combinations of acquisition parameter and 6-to-6 image pairs (\textbf{Supporting Information Figure S3}). The training process was repeated for all brain orientation views (axial, sagittal, and coronal) and for both the amide- and semi-solid MT-weighted data. Overall, a total of 563,904 image series/acquisition parameter pairs (3,383,424 single images) were used for core module development. 

The quantification module implements a transfer learning strategy, which benefited from, and expanded upon, the trained core module. Therefore, a relatively small dataset (85,824 image series and acquisition parameter pairs) was sufficient for its fine tuning.

All images were motion-corrected and registered using elastix\cite{klein2009elastix}. Skull removal was performed using statistical parameter mapping (SPM)\cite{ashburner2005unified} on a T$_1$ map. Quantitative reference semisolid MT-MRF maps were obtained using a fully connected neural network trained on simulated dictionaries, where all \textit{m}=30 raw input measurements were taken as input. Pixelwise T$_1$, T$_2$, and  B$_0$ values were also incorporated as direct NN inputs, for improved reconstruction accuracy\cite{perlman2022quantitative}. A detailed description of the ST-MRF reconstruction and quantification procedure has been published previously\cite{perlman2022quantitative, weigand2023accelerated}.

The core module 'label' images were derived from the physically acquired non-steady-state amide or semisolid MT data. The quantification module treated the separately acquired B$_0$, B$_1$, T$_1$, T$_2$, f$_{ss}$, and k$_{ssw}$ maps as the training set image labels.

The test cohort was composed of three separate datasets: (1) Two healthy volunteers scanned at TAU (not used for training or validation). (2) A healthy volunteer scanned at FAU using a scanner model and hardware different to those used for training. (3) A brain cancer patient scanned at FAU (see more details in the Human Subjects section above).

\subsection{Core module architecture.}
The core module (\textbf{Fig. 1a}) was designed to generate 'on-demand' image contrast, according to a user defined acquisition parameter set. It receives a dual-domain input, representing a per-subject (rapidly acquired) calibration image set and RF excitation information.  The calibration set was composed of serially acquired image data $x \in \mathbb{R}^{M \times H \times W}$, where M is the temporal dimension (6 in our case), and H $\times$ W are the spatial image dimensions. The RF information is represented by an acquisition parameter tensor $p \in \mathbb{R}^{2 \times (2M)}$, composed of the saturation pulse powers  ($B_1$) and frequency offsets ($\omega_{rf}$), associated with  the calibration and the 'on-demand' image sets, respectively. The module output is a set of new contrast images $y \in \mathbb{R}^{M \times H \times W}$. 

 Each calibration image was reshaped into 9 patches that were projected linearly and embedded into a tissue response representation. The acquisition parameter tensor was converted into RF excitation embedding, using a fully connected layer. The dual-domain embedding was then concatenated into a single tensor and transferred into a transformer encoder\cite{hatamizadeh2022unetr, dosovitskiy2020image}, with the following hyper-parameters: embedding dimension = 768, MLP size = 3072, transformer layers = 3, attention heads = 4.

The next step involved the sequential application of three convolution layer blocks. The first two blocks comprised 3x3 convolutions, batch normalization, ReLU activation function, and Max Pooling. The third block contained an up-sample layer, a $3\times$3 convolution layer, and a sigmoid activation function.

\subsection{Quantification module architecture.}
The quantification module was designed to leverage the intricate mapping between the RF irradiation domain and the image domain, as extensively learned by the core module, and then utilize a transfer learning strategy in order to achieve quantitative mapping. Specifically, the same weights used for the on-demand contrast generation task served as the initial state for the transformer encoder in the quantification module. The architecture (\textbf{Fig. 1b}) included several modifications: the last convolutional layer and sigmoid activation were replaced by a new 3$\times$3 convolutional layer, batch normalization, and ReLU activation. An additional (fourth) convolutional block was added, concluded by sigmoid activation. The quantification module input was identical to that of the core module, while the target output was six parameter maps: k$_{ssw}$, f$_{ss}$, B$_0$, B$_1$, T$_1$, T$_2$.

\subsection{Training Properties.}
For both modules, the loss function $\mathcal{L}$ was defined as a combination of the structural similarity index measure (SSIM) and L1.
The core module was trained using five RTX 5000 GPUs in parallel, using a batch size = 64, and learning rate = 0.0004. The training (259 epochs) took five days. The quantification module was trained using a single RTX 5000 GPU, with a batch size = 16, and learning rate = 0.002. The training (348 epochs) took three days. All models were implemented in PyTorch.

\subsection{Statistical analysis}
A two-tailed t-test was calculated using the open-source SciPy scientific computing library for Python\cite{virtanen2020scipy}. Differences were considered significant at P $<$ 0.05. In all box plots, the central 
horizontal lines represent median values, box limits represent the upper (third) and lower (first) quartiles, the whiskers represent 1.5 × the interquartile range above and  below the upper and lower quartiles, respectively, and outliers are presented as circles.

\subsection{Code availability}
The DeepMonC framework will be made publicly open upon acceptance at \\ https://github.com/momentum-laboratory/deepmonc. The full repository was uploaded as supporting information for the referees.

\subsection{Data availability}
The main data supporting the results of this study are available within the paper and Supplementary Information. The 3D human data cannot be shared due to subject confidentiality and privacy. Two sample 2D datasets will become available at https://github.com/momentum-laboratory/deepmonc upon acceptance.


\begin{figure}
\centerline{\includegraphics[height=3.5in,width=7.00in]{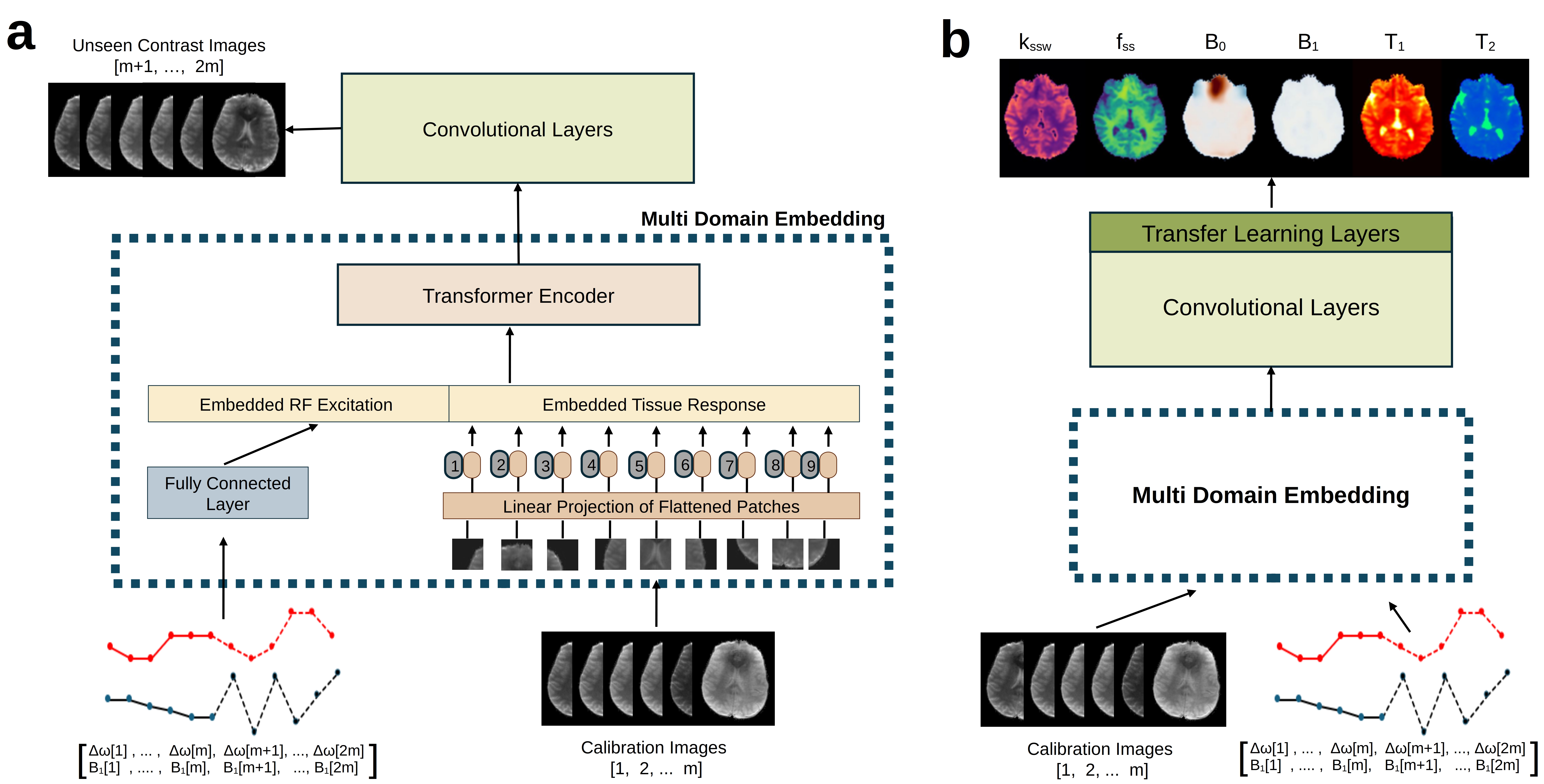}}
\caption{\textbf{Schematic representation of the biophysical-model-free deep MRI on a chip (DeepMonC) framework}. \textbf{a}. Automatic prediction of unseen molecular MRI contrast weighted images. A multi-domain input is used, including a sequence of m non-steady-state MRI calibration images and an RF excitation parameter tensor. It includes the acquisition parameters associated with the calibration images (solid lines) and the on-demand acquisition parameters (dashed lines) for the desired image output (m new images shown at the top). Separate embeddings for the real image data and the physical RF properties are learned using a vision transformer and a fully connected layer, respectively. \textbf{b}. A quantification module for the simultaneous mapping of six tissue and scanner parameter maps, including the semi-solid proton volume fraction (f$_{ss}$) and exchange rate (k$_{ssw}$), water proton longitudinal (T$_1$) and transverse (T$_2$) relaxation, and static (B$_0$) and transmit (B$_1$) magnetic fields. This module exploits the multi-domain embedding learned by the core module, utilizing a transfer learning strategy. }
\end{figure}

\begin{figure}
\centerline{\includegraphics[height=7.06in,width=6.50in]{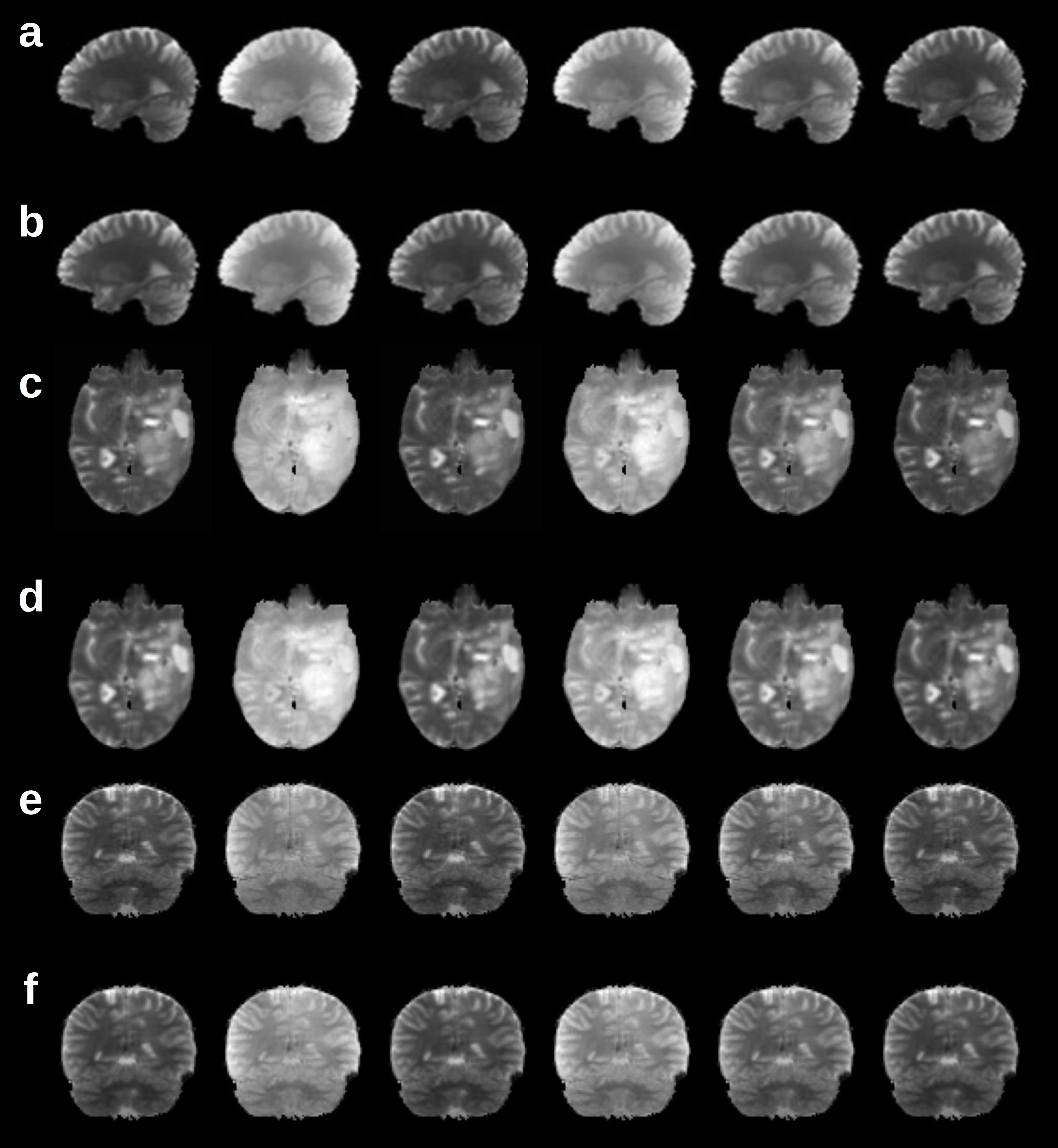}}
\caption{\textbf{Automatic prediction of unseen molecular MRI contrast weighted images}. A comparison between  representative ground truth \textbf{(a, c, e)} and DeepMonC-predicted \textbf{(b, d, f)} molecular MRI contrast-weighted images in the human brain. \textbf{(a, b)} Semiolid MT-weighted images from an unseen subject. \textbf{(c, d)} Amide proton transfer CEST-weighted images from a brain tumor patient scanned at an unseen imaging site. \textbf{(e, f)} Semisolid MT-weighted images from an unseen subject scanned at an unseen imaging site with hardware that was different from that used for training.}
\end{figure}

\begin{figure}
\centerline{\includegraphics[height=6in,width=6.00in]{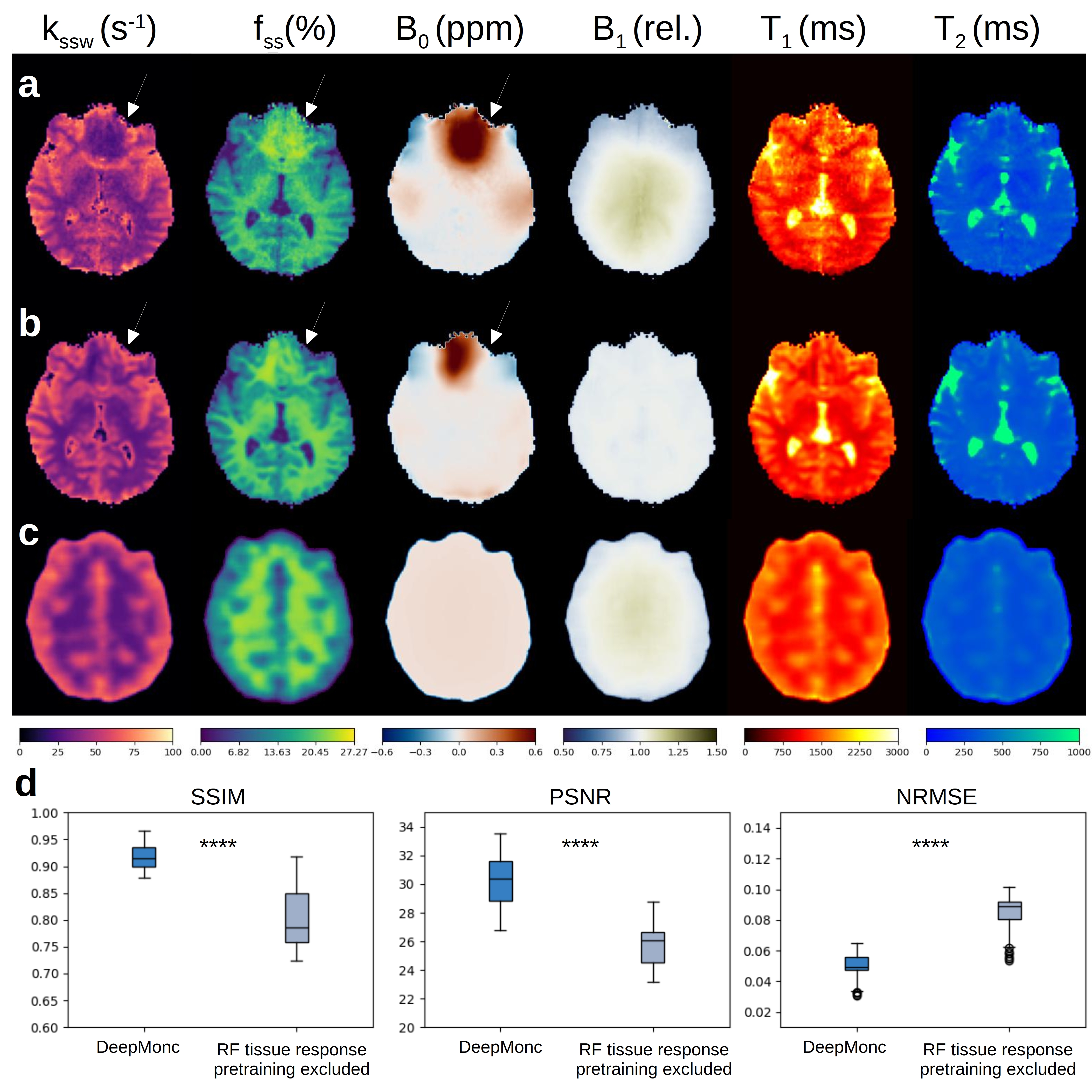}}
\caption{\textbf{Quantitative reconstruction of six molecular MRI, scanner field, and water-proton relaxation quantitative maps from a new healthy human volunteer scanned at the same imaging site used for training}. (\textbf{a}) Ground truth reference images obtained using conventional T$_1$ and T$_2$-mapping, WASABI, and semisolid MT MR-Fingerprinting (MRF) in 8.5 min. (\textbf{b}) The same parameter maps obtained using DeepMonC in merely 28.2 s (94\% scan time acceleration). Note the reduced field inhomogeneity (as seen in the B$_0$ and B$_1$ predicted images), which explains the successful noise reduction in the output maps (white arrows). (\textbf{c}) Quantitative reconstruction using conventional supervised learning (RF tissue response pretraining excluded), utilizing the same raw input data used in (b) for comparison. (\textbf{d}) Statistical analysis of the SSIM, PSNR, and NRMSE performance measures, comparing the DeepMonC reconstructed parameter maps to reference ground truth (n = 69 brain image slices per group ). ****p$<$0.0001.}
\end{figure}

\begin{figure}
\centerline{\includegraphics[height=6in,width=6.00in]{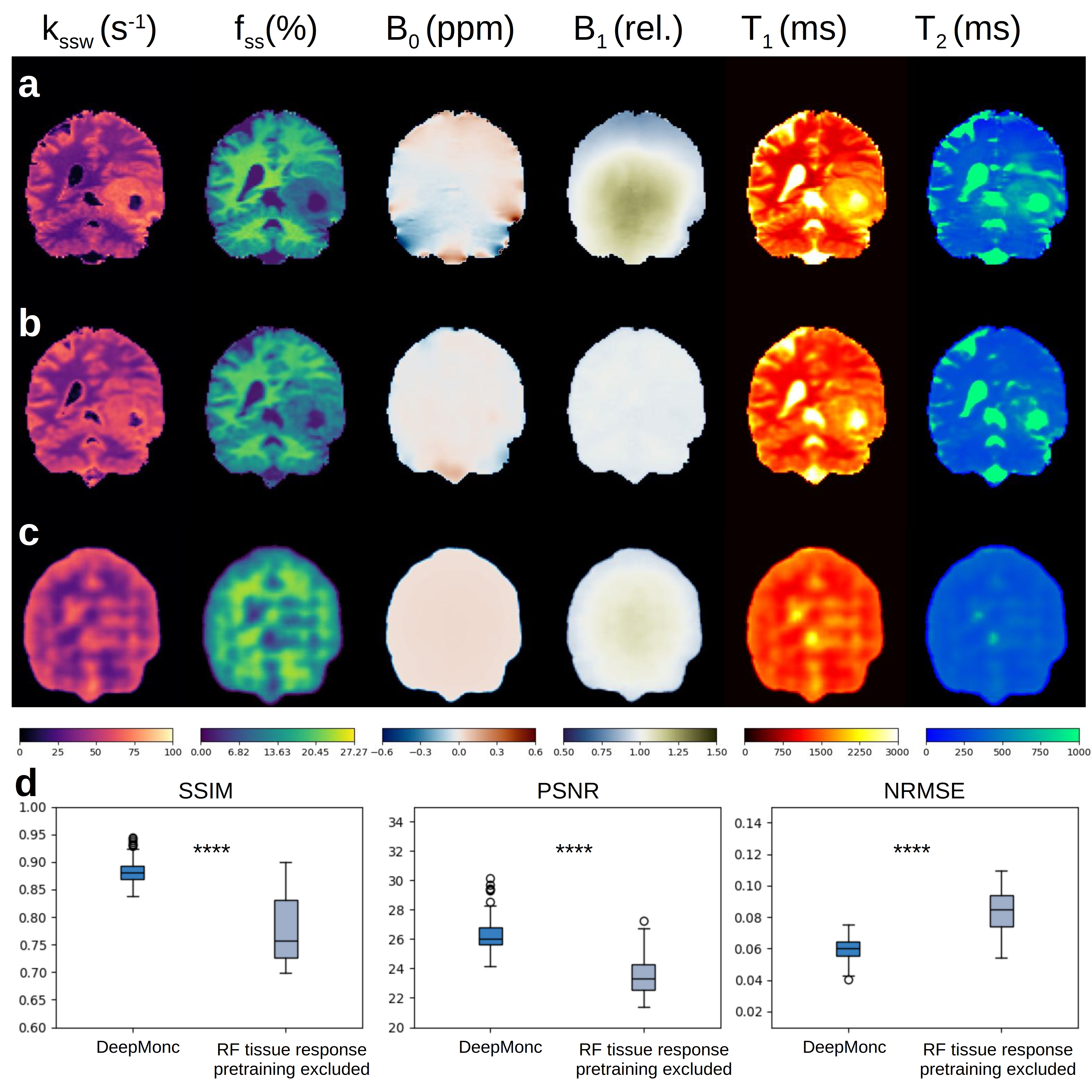}}
\caption{\textbf{Quantitative reconstruction of six molecular MRI, scanner field, and water-proton relaxation quantitative maps from a brain cancer patient scanned at a different imaging site compared to training}. (\textbf{a}) Ground truth reference images obtained using conventional T$_1$ and T$_2$-mapping, WASABI, and semisolid MT MR-Fingerprinting (MRF) in 8.5 min. (\textbf{b}) The same parameter maps obtained using DeepMonC in merely 28.2 s (94\% scan time acceleration). (\textbf{c}) Quantitative reconstruction using conventional supervised learning (RF tissue response pretraining excluded), utilizing the same raw input data used in (\textbf{b}) for comparison. (\textbf{d}). Statistical analysis of the SSIM, PSNR, and NRMSE performance measures, comparing the DeepMonC reconstructed parameter maps to reference ground truth (n = 68 brain image slices per group ). ****p$<$0.0001.}
\end{figure}

\begin{figure}
\centerline{\includegraphics[height=6in,width=6.00in]{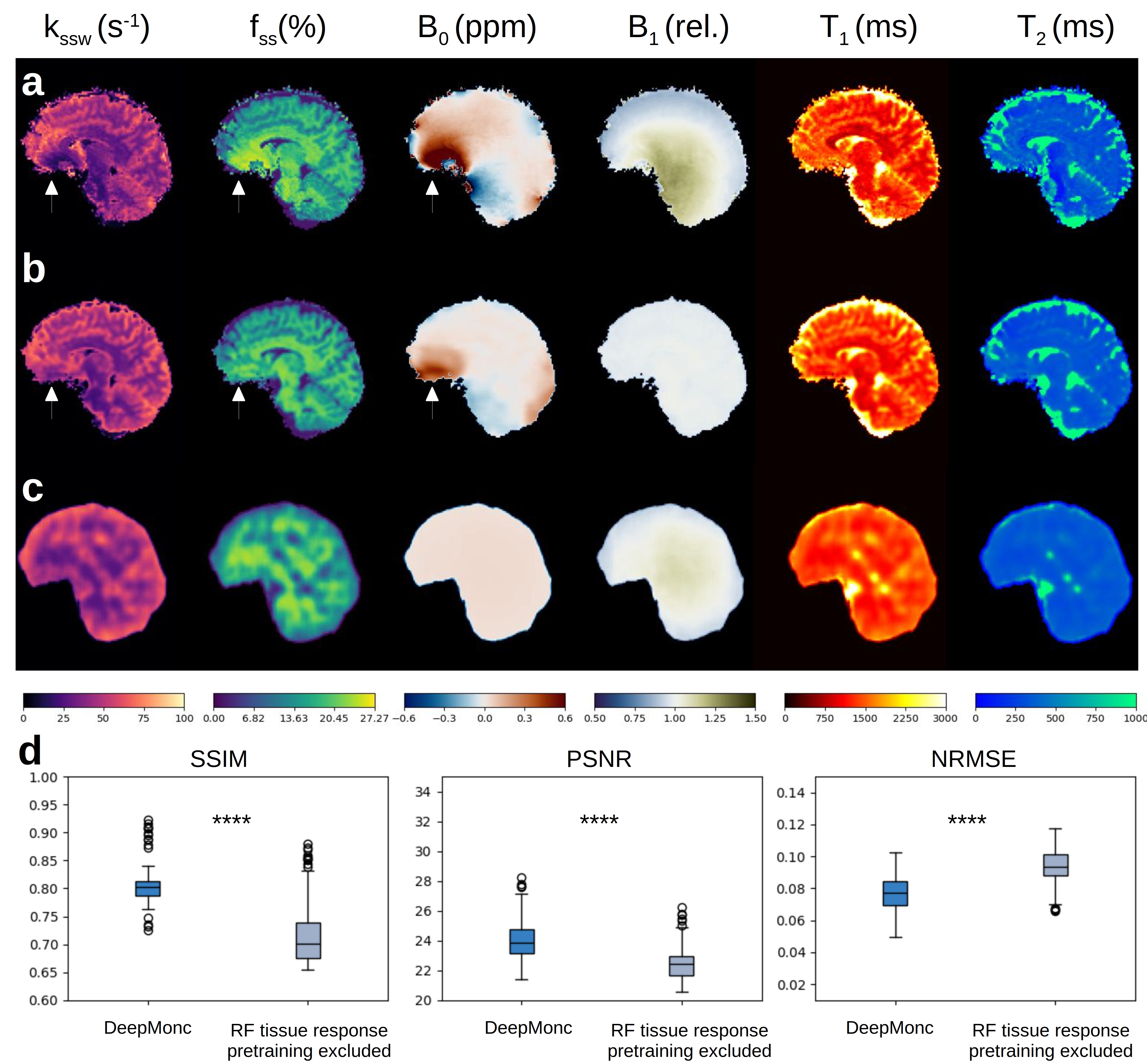}}
\caption{\textbf{Quantitative reconstruction of six molecular MRI, scanner field, and water-proton relaxation quantitative maps from a new healthy volunteer scanned at a different imaging site and different hardware compared to training}. (\textbf{a}) Ground truth reference images obtained using conventional T$_1$ and T$_2$-mapping, WASABI, and semisolid MT MR-Fingerprinting (MRF) in 8.5 min. (\textbf{b}) The same parameter maps obtained using DeepMonC in merely 28.2 s (94\% scan time acceleration). Note the reduced field inhomogeneity (as seen in the B$_0$ and B$_1$ predicted images), which explains the successful noise reduction in the output maps (white arrows). (\textbf{c}) Quantitative reconstruction using conventional supervised learning (RF tissue response pretraining excluded), utilizing the same raw input data used in (\textbf{b}) for comparison. (\textbf{d}) Statistical analysis of the SSIM, PSNR, and NRMSE performance measures, comparing the DeepMonC reconstructed parameter map to reference ground truth (n = 68 brain image slices per group ). ****p$<$0.0001.}
\end{figure}

\renewcommand{\figurename}{Table}
\setcounter{figure}{0}

\begin{figure}
\centerline{\includegraphics[height=3.35in,width=7.00in]{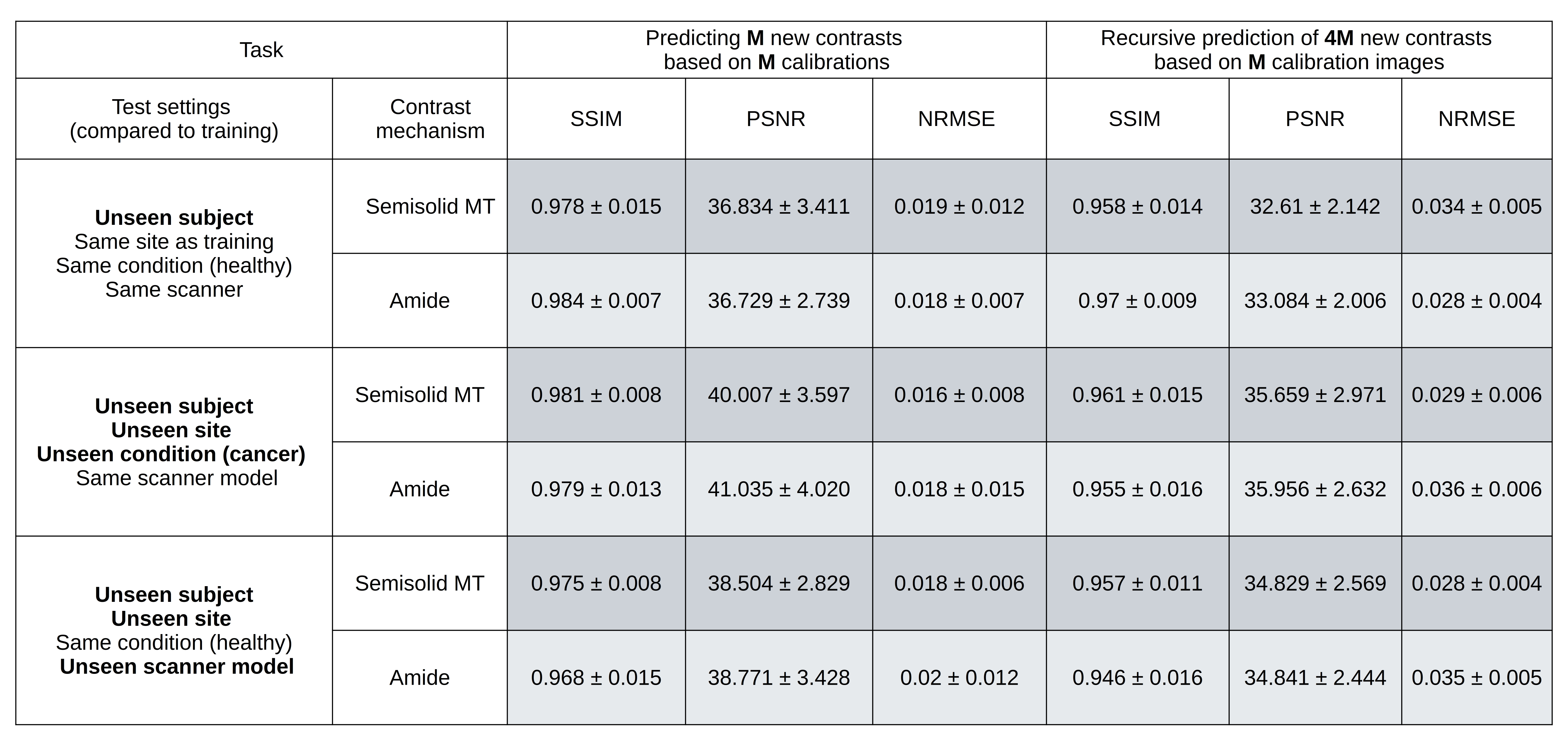}}
\caption{\textbf{Performance analysis for on-demand generation of molecular contrast-weighted images, comparing the DeepMonC reconstructed output to the reference ground truth.} 
SSIM - Structural similarity index measure; PSNR - peak signal-to-noise ratio; NRMSE - normalized mean-square error.}
\end{figure}

\end{document}